\documentclass[showpacs,preprintnumbers,amsmath,amssymb,floatfix,nofootinbib]{revtex4}

\usepackage{color}   
\usepackage{cancel}
\usepackage{color}
\usepackage{tikz}
\usetikzlibrary{decorations.pathmorphing,decorations.markings,trees}
\usepackage[utf8]{inputenc}
\usepackage[T1]{fontenc}
\usetikzlibrary{shapes,arrows,positioning,automata,backgrounds,calc,er,patterns}
\usepackage{tkz-graph}
\tikzstyle{vertex}=[circle, draw, inner sep=0pt, minimum size=6pt]

\usepackage{graphicx}
\usepackage{dcolumn}
\usepackage{float}
\usepackage{bm}

\usepackage[utf8]{inputenc}
\usepackage[T1]{fontenc}
\usepackage{ulem}
\usepackage{hyperref} 
\hypersetup{
    colorlinks=true,    
    urlcolor=blue,      
    linkcolor=red,      
    citecolor=green     
}

\begin{document}

\def\bb    #1{\hbox{\boldmath${#1}$}}
\def\bb    #1{\hbox{\boldmath${#1}$}}

\def\blambda{{\hbox{\boldmath $\lambda$}}}
\def\eeta{{\hbox{\boldmath $\eta$}}}
\def\bxi{{\hbox{\boldmath $\xi$}}}
\def\bzeta{{\hbox{\boldmath $\zeta$}}}
\def\sD{D \!\!\!\!/}
\def\sd{\partial \!\!\!\!/}
\def\EQ{{\hbox{\boldmath $Eq.(\ref$}}}
\def\qcd{{{}^{\rm QCD}}}   
\def\qed{{{}^{\rm QED}}}   
\def\2d{{{}_{\rm 2D}}}         
\def\4d{{{}_{\rm 4D}}}         

\today

\large

\begin{center}
  {\Large\bfseries Covariant Gross-Pitaevskii-like equation for relativistic fermions}\\[1em]
  Andrew Koshelkin\\
  {\small\itshape NRNU-MEPhI, Kasirskoye shosse, 31, 115409, Moscow, Russia}\\[0.6em]
\end{center}

\begin{abstract}
The covariant Gross-Pitaevskii-like equation for studying systems of interacting relativistic fermions is proposed. The Lagrangian generating  such an equation is derived. The considered equation is examined in  studying the fermi-liquid, superconductivity and the confinement problem in QCD. The Migdal's step in the fermi-liqiud model, the excitation spectrum  of superconductive fermion system are derived in the context of the proposed  Gross-Pitaevskii-like equation. We show that the quark fields governing such an equation turn out to be confined in the  (3+1) space time.
\end{abstract}


\maketitle

\vspace{0.8em}

\section{Introduction}

The Gross-Pitaevskii (GP) equation was proposed~\cite{Gro,Pit} to  study  interacting boson systems. Such an equation,  being phenomenological one in its nature, allows us eliminate real boson-boson interaction nonlinear terms and coupling constants. Argued by  the Hartree-Fock approximation  the obtained  equation successfully describes the Bose-Einstein condensate of interacting particles in terms of a single-particle wave function of a quasi-particle in the ground state. The derived  equation is the non-linear Schr$\ddot o$dinger equation in its structure, and has the following form in the time-dependent case.

\begin{eqnarray}\label{GP}
i \frac{ \partial \psi(\mathbf  r,t)}{\partial t}= \Bigl(  -\frac{1}{2m} \nabla^2 +V(\mathbf r) +g|\psi(\mathbf  r, t)|^2 \Big)\psi(\mathbf  r,t), 
\end{eqnarray}
where $\psi(\mathbf  r)$ is the wave function, $m$ is the particle (quasi-particle) mass,  $g$ is the coupling constant, $V(\mathbf r)$ is an external potential, $\hbar =c=1$.

In the framework of the simplest models the GP equation allows us to derive the exact solutions which describe observable effects. It, in  particularly,  concerns the Thomas-Fermi approach when the number of the condensate particle is large. Then, provided that the harmonic traps, the GP equation gives the correct profile of the density of condensate particles\cite{Foo}. The exact solution of the GR equation when $V(\mathbf r)=0$ supports the  Hugenholtz–Pines theorem\cite{Hug}. In the Bogolubov's approach\cite{Bog},  solving  the GP equation, the dispersion law  which  leads to the second sound is found\cite{Pit, Pit98}. The  GP equation in the  helical potential determines  the dynamics of optical vorteces\cite{Oku}.

Following the GP idea, we covariantly modify the standard Dirac equation  so that a new  equation  derived would govern a set of interacting relativistic fermion. There are many ways generalizing   the Dirac to   solve  the stated  problem. We take   the simplest  and the most natural way consisting in modification of  the mass term in the Dirac equation, making it to be  functionally depending on scalar products of  the Dirac and  Dirac-conjugate  fields to keep  the Lorentz invariance of a new equation. 
However, such an approach itself does not allow us to  solve a physical problem in the general case, since interaction in fermion system depends strongly on a lot various parameters characterizing the considered interacting fermions. Therefore, there is necessary to find such extensions of the Dirac equation which would be relevant  to describe this or that physical situation, and  leads to understandable results.  

The structure of the present paper is following. In the second section we derived modified the Dirac equation to describe interacting relativistic fermions, and write down the Lagrangian generating such an equation.  All  sections below will be devoted to examining the equation derived above. In the third section we consider the dynamics of fermions in the case of  small  coupling constants. Following the fermi-liquid approach we obtain the height of the Migdal step\cite{LL9} quasi-particles  in the case  the weak interacting fermions in the degenerated Fermi-liquid. The IV-th section is devoted to applying  the developed approach for  study the superconductivity of fermion systems. We obtain the value of the gap in the  energy spectrum of interacting fermions, the dispersion law  of a fermion system at low temperatures. The confinement in QCD is studied in Section V. A qaurk-antiquark pair is shown to be confined in the (3+1) Minkowskii space-time in the context of the developed approach.

\section{The generalized Dirac equation}

The standard Dirac equation contain the mass term  which is a true Lorentz scalar. We modify this term,   changing  a mass $m$   to keep the covariance  of the new Dirac-like   equation. The most natural  way to get to it is to assume that the effective mass depends also on  the products of type of  $\bar \psi (x)  \psi (x)$ and $\bar \psi (x) \gamma^5 \psi (x)$

\begin{eqnarray}\label{eq1}
m\to m_0 + {\frak g} {\frak F}[\bar \psi (x)  \psi (x), (\bar \psi (x)\gamma^5  \psi (x))^2], 
\end{eqnarray}
where $m_0$ is the mass of a real fermion, $ {\frak F}[\bar \psi (x)  \psi (x), (\bar \psi (x)\gamma^5  \psi (x))^2]$ is some differentiable  function, $  \psi (x)$ and $\bar \psi (x) $ are the Dirac and Dirac-conjugate bispinors,respectively, ${\frak g}$ is a coupling constant,  $\gamma^\mu, \mu =0,1,..., 5$ are the Dirac  $\gamma$-matrices; $x=(t,\mathbf r)$ denotes a point in the (3+1) Minkowskiii space-time. The function $ {\frak F}[\bar \psi (x)  \psi (x), (\bar \psi (x)\gamma^5  \psi (x))^2]$ introduced by such a way, takes into account phenomenologically interactions between fermions. 

Then, the Lagrangian  governing the dynamics such a fermion field is

\begin{eqnarray}
 \label{eq9}
&& {\cal L} = \frac{i}{2} \Big[\bar\psi (x) \Big( \gamma^\mu \partial_\mu +i(m_0 +  {\frak g} {\frak F}[\bar \psi (x)  \psi (x), (\bar \psi (x)\gamma^5  \psi (x))^2]\Big) \psi (x)\Big]-\nonumber \\
&& \frac{i}{2}  \Big[\bar\psi (x) \Big(\gamma^\mu {\overleftarrow \partial}_\mu - i(m_0 +  {\frak g} {\frak F}[\bar \psi (x)  \psi (x), (\bar \psi (x)\gamma^5  \psi (x))^2])\Big) \psi (x)\Big],
\end{eqnarray}

This Lagrangian generates the non-linear  Dirac equation

\begin{eqnarray}\label{eq2}
p_\mu \gamma^\mu  \psi (x) =(m_0 +g F[\bar \psi (x)  \psi (x), (\bar \psi (x)\gamma^5  \psi (x))^2]) \psi (x), 
\end{eqnarray}
where $\gamma^\mu$ are the Dirac matrices which will be taken  in the Weyl representation\cite{Pes95}, $g$ is the coupling constant,  $g_2$ is some coupling constant, $p_\mu = i\partial_\mu $. The function is defined by a formula

\begin{eqnarray}
 \label{eq10-1}
&& gF[\bar \psi (x)  \psi (x), (\bar \psi (x)\gamma^5  \psi (x))^2]= {\frak g} {\frak F}[\bar \psi (x)  \psi (x), (\bar \psi (x)\gamma^5  \psi (x))^2]+  {\frak g} \bar\psi (x)\frac{\delta {\frak F}[\bar \psi (x)  \psi (x), (\bar \psi (x)\gamma^5  \psi (x))^2]}{\delta \bar\psi (x)}
\nonumber \\
\end{eqnarray}
The derived equation is very close to the GP equation in its structure.

We note that according to the Fierz identities the following constraint relation particularly occurs

 \begin{subequations}
\begin{eqnarray}
 \label{Fierz}
&& j^\mu (x) j_\mu (x)  \equiv (\bar\psi (x)) \gamma^\mu (\psi (x) \bar\psi (x)) \gamma_\mu \psi (x)=
(\bar\psi (x) \psi (x))^2  -(\bar\psi (x) \gamma^5 \psi (x))^2 ;\\
&& j^\mu (x) j_\mu (x)  \equiv (\bar\psi (x) \tau_a \gamma^\mu \psi (x))( \bar\psi (x) \tau^a \gamma_\mu \psi (x))=\frac{N^2-1}{2N}\left((\bar\psi (x) \psi (x))^2  -(\bar\psi (x) \gamma^5 \psi (x))^2\right),
\end{eqnarray}
\end{subequations}
where $j^\mu (x)$ is a fermion current in the cases of $U(1)$ and $SU(N)$ symmetries, $\tau^a$ are the $SU(N)$ group generators\cite{Pes95}.
Then, the Lagrangian  (\ref{eq9}) can be written as follows

\begin{eqnarray}
 \label{eq9-1}
&& {\cal L} = \frac{i}{2} \Big[\bar\psi (x) \Big( \gamma^\mu (\partial_\mu - \frac{i }{2}g_1 \bar\psi (x) \gamma_\mu \psi (x))+i(m_0 +{\frak {\tilde g }}{\frak{ \tilde F}}[\bar \psi (x)  \psi (x), (\bar \psi (x)\gamma^5  \psi (x))^2]\Big) \psi (x)\Big]-\nonumber \\
&& \frac{i}{2}  \Big[\bar\psi (x) \Big(\gamma^\mu ({\overleftarrow \partial}_\mu +  \frac{i }{2}g_1 \bar\psi (x) \gamma_\mu \psi (x))- i(m_0 + {\frak {\tilde g }}{\frak{ \tilde F}}[\bar \psi (x)  \psi (x), (\bar \psi (x)\gamma^5  \psi (x))^2])\Big) \psi (x)\Big],
\end{eqnarray}
so that

\begin{eqnarray}
 \label{eq9-2}
&&{\frak {\tilde g }} {\frak{ \tilde F}}[\bar \psi (x)  \psi (x), (\bar \psi (x)\gamma^5  \psi (x))^2]={\frak g} {\frak{ F}}[\bar \psi (x)  \psi (x), (\bar \psi (x)\gamma^5  \psi (x))^2] +\nonumber \\
&&\frac{g_1}{2}\left( (\bar\psi (x) \psi (x))^2  -(\bar\psi (x) \gamma^5 \psi (x))^2\right), 
\end{eqnarray} 
where $g_1$ some coupling constant.

Then, the nonlinear Dirac equation can be formally  present in the form of the Dirac equation in an external gauge field, where the role of the gauge field $A_\mu (x) $ plays the current $j_\mu (x) $

\begin{eqnarray}\label{eq8}
(p_\mu \gamma^\mu +g_1 j_\mu (x) \gamma^\mu )  \psi (x) =(m_0 + { { g_2 }} f[\bar \psi (x)  \psi (x), (\bar \psi (x)\gamma^5  \psi (x))^2]) \psi (x). 
\end{eqnarray}
where $ f[\bar \psi (x)  \psi (x), (\bar \psi (x)\gamma^5  \psi (x))^2])$ is derived from $ {\frak{\tilde F}}[\bar \psi (x)  \psi (x), (\bar \psi (x)\gamma^5  \psi (x))^2])$, using Eqs.(\ref{eq10-1}), (\ref{eq9-2}).

The  equation governing dynamics of  $\bar\psi (x)$ can be written in one of the following form

\begin{eqnarray}\label{eq11}
- p_\mu \gamma^\mu  \bar\psi (x) =(m_0 +g F[\bar \psi (x)  \psi (x), (\bar \psi (x)\gamma^5  \psi (x))^2] ) \bar\psi (x). 
\end{eqnarray}

\begin{eqnarray}\label{eq11-1}
- (p_\mu \gamma^\mu - g_1 j_\mu (x) \gamma^\mu )  \bar\psi (x) =(m_0 +{ { g_2 }} f[\bar \psi (x)  \psi (x), (\bar \psi (x)\gamma^5  \psi (x))^2] ) \bar\psi (x). 
\end{eqnarray}

Further, we examine the derived GP-like equation in various cases to describe physical phenomena taking place in interacting fermion systems .
 
 \section{ The non-liner Dirac equation under  weak interaction}
 
 We consider Eqs.(\ref{eq8}),  (\ref{eq11}) when the coupling constants $g_1$ and $g_2$ are small\footnote{Since the introduced coupling constants are dimensional, the inequalities defining them smallness is managed to write for each considered problem only rather than in the  general case .}. For brevity,  we consider consider OED situation when a single current occurs. in the leading approximation we omit all terms containing the coupling constants. As a result, we go to the free Dirac equation which solution is\cite{Pes95}

\begin{eqnarray}
 \label{eq12}
&& \psi_0 (x) =\sum_s \int \frac{d^3\bb p}{\sqrt{2 \varepsilon (\bb p)}(2\pi)^3}  \left( a_s (\bb p)  u_s (\varepsilon (\bb p), \bb p)\exp{(-ipx)} +b^\dag_s (\bb p)  v_s (\varepsilon (\bb p), -\bb p)\exp{(+ipx)}\right)
\end{eqnarray} 
This solution can be rewritten as follows
\begin{eqnarray}\label{eq12-1}
&&\psi_0 (x)  = \int  \frac{d^4 p}{(2\pi)^4} \psi_0 (p) \exp{(-ipx)} =\nonumber \\
&&\sum_s \int\limits_{-\infty}^{+\infty} \frac{id\omega }{(2\pi)}\int \frac{d^3\bb p}{\sqrt{2 |\omega|}(2\pi)^3}  \left( \frac{ a_s (\bb p)  u_s (\varepsilon (\bb p), \bb p)\exp{(-ipx)}}{\omega - \varepsilon (\bb p)+i\delta \operatorname{sgn}(\omega)}-  \frac{b^\dag_s (-\bb p)  v_s (\varepsilon (\bb p), \bb p)\exp{(-ipx)}}{\omega + \varepsilon (\bb p)+i\delta \operatorname{sgn}(\omega)}\right)\nonumber \\
\end{eqnarray} 
where $p= (\omega, \bb p)$ is the  4-momentum of a particle (antiparticle); $\varepsilon (\bb p)= +\sqrt{\bb p^2 + m_0^2}$ is the fermion (antifermion) energy; $a_s (\bb p)$ and $a^\dag_s (\bb p) $ are the operators of annihilation and creation of a fermion, whereas $b_s (\bb p)$ and $b^\dag_s (\bb p) $ are the operators of annihilation and creation of an antifermion, $ u_s (\varepsilon (\bb p), \bb p)$ and $ v_s (\varepsilon (\bb p), \bb p)$ are the Dirac bispinors in the Weyl representation\cite{Pes95}. The solution (\ref{eq12}) is normalized number of particles $n_0$ per unit volume.

\begin{eqnarray}
 \label{eq13}
&& n_0=<:\psi^\dag (x)\psi(x):> =\sum_s \int \frac{d^3\bb p}{(2\pi)^3}  \left(< a^\dag_s (\bb p) a_s (\bb p)> +<b^\dag_s (\bb p) b_s (\bb p)>  \right) = \nonumber \\
&&\sum_s \int\limits_{-\infty}^{+\infty} \frac{id\omega }{(2\pi)}\int \frac{d^3\bb p}{ (2\pi)^3}  \left( \frac{ n_s (\bb p) }{\omega - \varepsilon (\bb p)+i\delta \operatorname{sgn}(\omega)}-  \frac{{\bar n}_s (\bb p) }{\omega + \varepsilon (\bb p)+i\delta \operatorname{sgn}(\omega)}\right),
\end{eqnarray} 
where $n_s (\bb p)$ and ${\bar n}_s (\bb p)$ are the occupancy numbers of particles and antiparticles, respectively.

Let us to derive corrections to the solutions (\ref{eq12}), restricting ourselves by the terms which linear with respect to the coupling constants. We substitute $\psi (x)$ in Eq.(\ref{eq2}) in the following form

\begin{eqnarray}
 \label{eq14}
&&\psi (x) = \psi_0(x) +\delta \psi (x) 
\end{eqnarray} 
After that, going to the momentum representation and keeping  the leading terms with respect to the coupling constant  $g$, we obtain (see Appendix A)

\begin{eqnarray}
 \label{eq15}
&&\delta \psi (p) = \frac{2m_0 gF [\bar \psi_0 \psi_0]}{p^2 - m_0^2} \psi_0(p),
\end{eqnarray} 
where $p=(\omega, \mathbf p)$. The function $ F[\bar \psi_0 \psi_0]$ depends only on $\bar \psi_0 \psi_0$ since $\bar \psi_0 \gamma^5 \psi_0 =0$. Moreover, since the coupling constant is small we have approximate changed $\bar \psi_0 \psi_0 \to  <:\bar \psi_0 \psi_0:> = constant $. 

Substituting the solution (\ref{eq15}) into Eq.(\ref{eq12-1}) and carrying out the integration needed, taking into account the decomposition given by  Eq.(\ref{a4}), we derive 

 \begin{eqnarray}\label{eq15-2}
&&\delta \psi (x)  =-\sum_s \int\limits_{-\infty}^{+\infty} \frac{id\omega }{(2\pi)}\int \frac{d^3\bb p}{\sqrt{2 |\omega|}(2\pi)^3} \frac{m_0 gF [\bar \psi_0 \psi_0]}{\omega^2}\nonumber \\
&& \left( \frac{ a_s (\bb p)  u_s (\varepsilon (\bb p), \bb p)\exp{(-ipx)}}{\omega - \varepsilon (\bb p)+i\delta \operatorname{sgn}(\omega)}-  \frac{b^\dag_s (-\bb p)  v_s (\varepsilon (\bb p), \bb p)\exp{(-ipx)}}{\omega + \varepsilon (\bb p)+i\delta \operatorname{sgn}(\omega)}\right)\nonumber \\
\end{eqnarray} 

Substituting $\psi(x) = \psi_0 (x) +\delta\psi (x) $ into Eq.(\ref{eq13}),  we derive

\begin{eqnarray}
 \label{eq16}
&& n_0=\sum_s \int\limits_{-\infty}^{+\infty} \frac{id\omega }{(2\pi)}\int \frac{d^3\bb p}{ (2\pi)^3}  \Bigg[ \frac{ n_s (\bb p) }{\omega - \varepsilon (\bb p)+i\delta\operatorname{sgn}(\omega)}\left( 1 -\frac{4m_0 gF [\bar \psi_0 \psi_0]}{\omega^2}\right)- \nonumber \\
&& \frac{{\bar n}_s (\bb p) }{\omega + \varepsilon (\bb p)+i\delta \operatorname{sgn}(\omega)}\left( 1 -\frac{4m_0 gF [\bar \psi_0 \psi_0]}{\omega^2}\right)\Bigg].
\end{eqnarray}

We analyses Eq.(\ref{eq16}) in the fermi-liquid approach\cite{LL9} for studying the set of nonrelativistic interacting  fermions.  We  consider such fermions at  the temperature  $T=0$.  We note that the terms in the square bracket in Eq.(\ref{eq16}) are the Green's functions of the fermions\cite{LL9}. In the fermi-liquid approach the only signature of the Green's function is the zero of the denominators in Eq.(\ref{eq16}). Besides that  ${\bar n}_s (\bb p)= { n}_s (-\bb p) $ which is the regular function around fermi-surface at $T=0$, and is equal to $ { n}_s (-\bb p_F)={ n}_s (\bb p_F)\simeq 1 $, where $p_F$ is the fermi-momentum.
Changing\cite{LL9} $\varepsilon (\bb p) \to \varepsilon (\bb p) - \varepsilon (\bb p_F)\sim v_F (p-p_F)$, we derive from Eq.(\ref{eq16})

\begin{eqnarray}
 \label{eq16-1}
&& n_0=\sum_s \int\limits_{-\infty}^{+\infty} \frac{id\omega }{(2\pi)}\int \frac{d^3\bb p}{ (2\pi)^3}  \Bigg[ \frac{ \left( 1 -\frac{4m_0 gF [\bar \psi_0 \psi_0]}{m_0^2}\right)}{\omega - v_F (p-p_F)+i\delta\operatorname{sgn}((p-p_F))}- \nonumber \\
&& \frac{\left( 1 -\frac{4m_0 gF [\bar \psi_0 \psi_0]}{m_0^2}\right) }{\omega +v_F (p-p_F)+i\delta\operatorname{sgn}((p-p_F))}\Bigg].
\end{eqnarray} 
where $v_F$ is the Fermi velocity.

Comparing Eq.(\ref{eq16-1}) with the formula (10.3)\cite{LL9}, we obtain for the height of  the  Migdal's step\cite{LL9}

\begin{eqnarray}
 \label{eq17}
&&H_M =   1- \frac{4 gF [\bar \psi_0 \psi_0]}{m_0}.
\end{eqnarray} 

In this case of non-relativistic particles    $\bar \psi_0 \psi_0 =n_0=p^3_F /3\pi^2$. Let us take  $F [\bar \psi_0 \psi_0]$ to be

\begin{eqnarray}\label{eq17-1}
 F [\bar \psi_0 \psi_0]= m_0 (\bar \psi_0 \psi_0)^{2/3}.  
\end{eqnarray}
Then, we have

\begin{eqnarray}
 \label{eq18}
&&H_M =   1- \frac{4g p_F^2}{9^{1/3} \pi^{4/3}}
\end{eqnarray}

Comparing with the classical  result, given by Eq.(22.21)\cite{LL9}, we find that

\begin{eqnarray}
 \label{eq18-1}
&& g= \frac{ a^2  9^{1/3} 2 \ln 2}{\pi^{2/3}},
\end{eqnarray}
where  $a$ is the scattering length.

We note, that the derived coupling constant depends  on the only  characteristics determining interaction between fermions which is the scattering length. Such a dependence of the coupling constant on the fermi-liquid parameters has dictated choosing the explicit form of an extension model which is given by Eq.(\ref{eq17-1}). The application limit of the developed consideration is $g \ll p_F^{-1}$.

\section{The superconductivity and  non-linear Dirac equation and}

As it has been done in the section above, we restrict our consideration by the QED case. Let us  go from Eq.(\ref{eq8}) to the second order equation by the standard way\cite{Pes95,LL4}. As a result, we obtain

\begin{eqnarray}\label{eq20}
(p^\mu  +g_1 j^\mu (x)  )^2  \psi (x) =(m_0 +  g_2 f[\bar \psi (x)  \psi (x)] )^2 \psi (x). 
\end{eqnarray}

 Assuming that the convolutions of type of $\bar \psi (x)  \psi (x)\simeq \bar \psi_0   \psi_ 0$, which are   in both  the $ j^\mu (x)$-term and  right-hand side of Eq.(\ref{eq8}),  are independent   on $x$, we go to the momentum representation in Eq.(\ref{eq20}). As a result, we obtain
 
\begin{eqnarray}\label{eq22}
(p^0 +g_1 j^0_{(0)}  )^2 \psi (p) =[(m_0 +   f[\bar \psi_0   \psi_0 ] )^2 + (\mathbf p +g_1 \mathbf j )^2]\psi (p). 
\end{eqnarray}
where the subscriber only means that $j^0_{(0)} $ is independent on $x$.

Let us go to the non-relativistic limit in Eq.(\ref{eq22}), that means that $\bar \psi_0   \psi_0 =  \psi^\dag_0   \psi_0 \simeq <:  \psi^\dag_0   \psi_0 :> =n_0$, where $n_0$ is the number of particles per unit volume. In the non-relativistic case we have $p^0 = m_0 + \varepsilon (\mathbf p)$, where $\varepsilon (\mathbf p)$ is the quasi particle energy\cite{LL9} . Then, taking the coupling constant to be 

\begin{eqnarray}\label{eq23}
g_1= -\frac{m_0}{n_0}, 
\end{eqnarray}
we derive from  Eq.(\ref{eq22}).

\begin{eqnarray}\label{eq24}
\varepsilon^2 (\mathbf p )\psi (p) =\left[(m_0 +g_2   f[\bar \psi_0   \psi_0 ] )^2 + \left(\mathbf p - \frac{m_0}{n_0}  \mathbf j \right)^2\right]\psi (p). 
\end{eqnarray}

Further, we pick $  f[\bar \psi_0   \psi_0 ]$ as follows\footnote{Such a choose of the extension model of the Dirac equation will be clear below.}

\begin{eqnarray}\label{eq25}
g_2  f[\bar \psi_0   \psi_0 ] = g_1( \bar \psi_0   \psi_0) + g ( \bar \psi_0   \psi_0)^{1/2} /v^2_F. 
\end{eqnarray}

We substitute $f[\bar \psi_0   \psi_0 ]$, given by the equation above, into  Eq.(\ref{eq24}), and consider this equation around the Fermi surface of fermion system. Then, we derive

\begin{eqnarray}\label{eq26}
v_F^2 \varepsilon^2 (\mathbf p ) =\left[(g n_0 )^2 +v_F^2 \left(\mathbf p - \mathbf p_F \right)^2\right], 
\end{eqnarray}
where $p_F$ is the Fermi momentum. In obtaining  Eq.(\ref{eq26}) we take into account that $ \mathbf j = n_0 \mathbf v_F$ around the Fermi surface.

We note that $ \varepsilon (\mathbf p )$ is the energy of a particle in the momentum space rather that the excitation energy ${\frak e }(\mathbf p )$ of the particle, which  is on the Fermi surface,  and considered in the energy space, as it takes place  are in the BCS superconductivity theory\cite{Bar}. To go from $ \varepsilon (\mathbf p )$ to ${\frak e }(\mathbf p )$ we have to conserve the total energy $ E$ of  fermion system, demanding 

\begin{eqnarray}\label{eq27}
E = \int \varepsilon (\mathbf p ) d N(\mathbf p )= \int {\frak e }(\mathbf p ) d N({\frak e })
\end{eqnarray}

On other hand, the last relation  means 

\begin{eqnarray}\label{eq28}
  d N({\frak e }) = d N(\mathbf p )\frac{dp}{d \varepsilon (\mathbf p )},
\end{eqnarray}
that directly  leads to
\begin{eqnarray}\label{eq29}
  \varepsilon (\mathbf p ) =\frac{dp}{d \varepsilon (\mathbf p )} {\frak e }(\mathbf p ). 
\end{eqnarray}

Since $\varepsilon (\mathbf p ) =p^2/2m$, we get

\begin{eqnarray}\label{eq30}
v_F  \varepsilon (\mathbf p ) = {\frak e }(\mathbf p ) , 
\end{eqnarray}
where ${d \varepsilon (\mathbf p )}$ is calculated on the Fermi surface. As a result,  we obtain the standard superconductivity dispersion law\cite{Bar}

\begin{eqnarray}\label{eq31}
{\frak e }(\mathbf p ) =\left[\Delta^2 +v_F^2 \left(\mathbf p - \mathbf p_F \right)^2\right], 
\end{eqnarray}
where the gap is $\Delta = (|g| n_0 )$.

We note such a dependence of the gap on the problem parameters is in a good relation to the BCS model 
(see, for example, Eq.(39.12) in Ref.\cite{LL9}), that has determined choosing the extension model of the Dirac equation in the form given by Eq.(\ref{eq25}).

\section{Quark confinement in the framework of  non-linear Dirac equation}

The color confinement of quarks in the standard (3+1) QCD is a very important problem which mathematically exact solution is very complicate. We apply the K.Wilson idea\cite{Wil} to study this problem in the context of the derived nonlinear Dirac equation. Following\cite{Wil} the probability $P$ of the confinement is 

\begin{eqnarray}\label{eq32}
W =\left< \exp\left(i \frac{N^2-1}{2}g_1\oint\limits_{C} dx^\mu j_\mu (x ) \right)\right>,
\end{eqnarray}
$ j_\mu (x )=Tr \{\tau_a \tau^b \} \bar \psi (x)\gamma_\mu \psi (x)$ is a quark current, the integration contours  $C$  is pictured on Fig.1.

 \begin{tikzpicture}[node distance=1cm and 1.5cm]
\tikzset{
particle/.style={solid,draw=black, line width=1.0pt, decoration={markings, mark=at position 0.57 with {\arrow[very thick]{<}}}, postaction={decorate}    },
gluon/.style={  decorate, draw=black,    decoration={coil,aspect=0}}
 }

\tikzset{
gluon/.style={  decorate, draw=black,    decoration={coil,aspect=0}}}

\tikzset{
gluon/.style={  decorate, draw=black,    decoration={coil,aspect=0.5}}
 }

\begin{tikzpicture}
  \draw[thick] plot [smooth cycle] coordinates 
    {(12,0) (8,0.5) (9,2) (9.5,3) (12.5,5)};
\end{tikzpicture}

\draw[->, thick] (-2.2,4.69) -- (-1.8,4.91);
\draw[->, thick] (-2.2,0.39) -- (-2.8,0.49);

\draw (-2.3,5.2) node [above]{$C$};

\draw (-5.5,0.4) node [above]{$(0,0)$};
\draw (-0.8,6.0) node [above]{$(T/2, \mathbf x)$};

\filldraw(-4.9, 1.2) circle[radius=1.7pt];
\filldraw(-0.4, 5.7) circle[radius=1.7pt];
\end{tikzpicture}

\begin{figure}[ht]
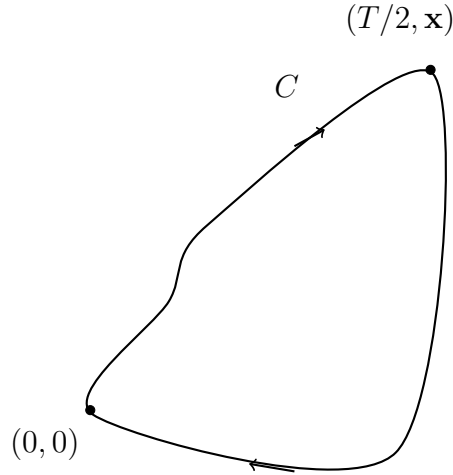

   \caption{The integration contour. }
\end{figure}

The  Wilson' loop is explicitly written as follows

\begin{eqnarray}\label{eq33}
W = \frac{\int {\cal D }\bar \psi {\cal D } \psi \exp\left(i g_1 \frac{N^2-1}{2}\oint\limits_{C} dx^\mu j_\mu (x ) +i \int d^4x L \right)}{\int{\cal D} \bar \psi{\cal D } \psi \exp\left(i \int d^4x L \right)},
\end{eqnarray}
where the Lagrangian $L$ is given by Eq.(\ref{eq9-1}).

Following A.Polyakov\cite{Pol}, we consider very narrow loop with respect to the space coordinate, going $T\to \infty (T\gg V^{1/3}\gg1 )$, where $V^{1/3}$ is the 3D volume  in the (3+1) space-time which is confined by the loop pictured in Fig.2. 

 \begin{tikzpicture}[node distance=1cm and 1.5cm]
\tikzset{
particle/.style={solid,draw=black, line width=1.0pt, decoration={markings, mark=at position 0.57 with {\arrow[very thick]{<}}}, postaction={decorate}    },
gluon/.style={  decorate, draw=black,    decoration={coil,aspect=0}}
 }

\tikzset{
gluon/.style={  decorate, draw=black,    decoration={coil,aspect=0}}}

\tikzset{
gluon/.style={  decorate, draw=black,    decoration={coil,aspect=0.5}}
 }

\begin{tikzpicture}
  \draw[thick] plot [smooth cycle] coordinates 
    {(9,0) (8.7,0.5) (9,1.2) (9.5,1.7) (12.5,5)};
\end{tikzpicture}

\draw[->, thick] (-2.2,3.4) -- (-1.7,3.91);
\draw[->, thick] (-2.2,2.5) -- (-2.8,1.7);

\draw (-2.3,5.2) node [above]{$C$};

\draw (-4.5,0.0) node [above]{$(0,0)$};
\draw (-0.5,6.0) node [above]{$(T/2, \mathbf x)$};

\filldraw(-3.85, 0.4) circle[radius=1.7pt];
\filldraw(-0.1, 5.6) circle[radius=1.7pt];
\end{tikzpicture}

\begin{figure}[ht]
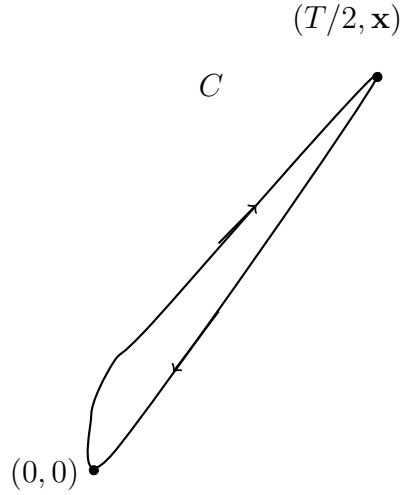

   \caption{The Polykov's integration contour. }
\end{figure}

here $n_0= <j^0> $ is a number of particle per unit 3D volume.

To calculate the function integrals in Eq,(\ref{eq33}) we use the functional determinant technique (see, for example,\cite{Pes95}). We  firstly note that

\begin{eqnarray}\label{eq36}
\oint\limits_{C} dx^\mu j_\mu (x ) = \int d^4 x \frac{\bar \psi \gamma^0 \psi}{V}=n_0 \int (\bar \psi \gamma^0  \psi)d^4 x , 
\end{eqnarray}
where $n_0= <j^0> $ is a number of particle per unit 3D volume.

Then, we have

\begin{eqnarray}\label{eq37}
W = \frac{\int {\cal D }\bar \psi {\cal D } \psi \exp\left(\int d^4x  \left(i g_1 n_0\frac{N^2-1}{2}  (\bar \psi \gamma^0  \psi) +i  L \right) \right)}{\int{\cal D} \bar \psi{\cal D } \psi \exp\left(i \int d^4x L \right)}= \det \left( 1+\frac {i g_1 n_0 \gamma^0\frac{N^2-1}{2} }{L}\right),
\end{eqnarray}
Such a determinant is\cite{Pes95}

\begin{eqnarray}\label{eq38}
 &&\det \left( 1+\frac {ig_1 n_0 \gamma^0 }{L}\frac{N^2-1}{2}\right)=\exp\left\{\frac{N^2-1}{2}\sum\limits_{n=1}^\infty \frac{1}{n} \left(-Tr\frac{i g_1 n_0 \gamma^0}{L}\right)^n \right\}=\nonumber \\
 &&\exp\left\{\frac{N^2-1}{2}\sum\limits_{n=1}^\infty \frac{1}{n}\left( -Tr\frac{ig_1 n_0 \gamma^0}{i\partial_\nu \gamma^\nu + g_1j_\nu \gamma^\nu - m_0) }\right)^n \right\},
\end{eqnarray}
Provided that the eigenvectors of the Lagrangian are known, we calculate the trace with respect such a eigenvalue set. For a brevity, we introduce  the momentum $\pi_\mu =p_\mu +g_1 j_\mu= i\partial_\mu +g_1 j_\mu$. Then,

\begin{eqnarray}\label{eq39}
 &&Tr\left(\frac{\gamma^0}{i\partial_\nu \gamma^\nu + g_1j_\nu \gamma^\nu - m_0) }\right)^n =Tr\left(\frac{-\pi_0 - 2i g_1j_0 -m_0\gamma^0}{-\pi^2 - m^2_0 }\right)^n
\end{eqnarray}

Since the integration contour strongly extended along $x^0$ axis $|\pi_0 |= |p^0|\gg \max (|\mathbf \pi| m_0) $,  we have  in the developed approach

\begin{eqnarray}\label{eq40}
 &&Tr\left(\frac{-\pi_0 - 2i g_1j_0 -m_0\gamma^0}{-\pi^2 - m^2_0 }\right)^n =4 (1/|p^0 |)^n =4(T)^n
\end{eqnarray}

Substituting Eq.(\ref{eq40}) into  Eq.(\ref{eq38}),  and going to the Euclidean space\cite{Wil,Pol} according to a formula $T\to iT$,  we obtain

\begin{eqnarray}\label{eq41}
W =  \exp\left\{\frac{N^2-1}{2}\sum\limits_{n=1}^\infty \frac{1}{n}\left( - 4 |g_1| n_0T\right)^n \right\}
\end{eqnarray}
The series in the exponent is\cite{Gra}

\begin{eqnarray}\label{eq41-1}
\sum\limits_{n=1}^\infty \frac{1}{n}\left( - 4 |g_1| n_0T\right)^n = -\ln(1+4|g_1| n_0 T), 
\end{eqnarray}
where we assume that attraction between fermions occurs, i.e. $g_1< 0$.
Then, we obtain for the probability

\begin{eqnarray}\label{eq42}
W =\frac{1}{(1+4|g_1| n_0 T)^{\frac{N^2-1}{2}}}
\end{eqnarray}

The obtained results support the  confinement, since at  fixed $g_1$ and $n_0$ and large $T$, the probability tends to zero. Moreover, , the more color number, the stronger confinement. We note if we, formally,  extrapolate Eq.(\ref{eq42}) to the case $N=1$, then this equation will be correct in the QED case, where no confinement at all in the (3+1) space-time.

\section{Conclusion}

The nonlinear GP-like Dirac equation is proposed for describing interaction fermions where real interaction between particle is phonologically included by means of nonlinear terms. The derived equation is covariant in the (3+1) space-time as compared, for example, the two-body relativistic problem.  The obtained equation is examined by applying it for to study observable physical phenomena in the framework of good tested model, such as the fermi-liquid model, the BCS superconductivity model, the confinement problem. The relevant  choosing the extension model of the Dirac equation is shown to lead to   good relations to the well-known results obtained  by the other authors earlier.

\appendix{}

\section{Solution of the nonlinear Dirac equation at small coupling constant. }

We look for the solution of Eq.(\ref{eq2})  at the  small coupling constant $g$ in the form

\begin{eqnarray}
 \label{a1}
&&\psi (x) = \psi_0(x) +\delta \psi (x),
\end{eqnarray}
where $\psi_0(x)$ is the the solution of the free Dirac equation given by  Eq.(\ref{eq12}).

Assuming that $\bar \psi_0 (x)  \psi_0 (x) \simeq constant$ we obtain in the momentum representation

\begin{eqnarray}\label{a2}
p_\mu \gamma^\mu (\psi_0(p) +\delta \psi (p)) =(m_0 +g F[\bar \psi_0   \psi_0]) (\psi_0(p) +\delta \psi (p)), 
\end{eqnarray}
where we take into account that  $ (\bar \psi_0 (x)\gamma^5  \psi_0 (x))=0$.

Holding the leading terms with  respect to small  $g$ in  Eq.(\ref{a1}) we derive from  Eq.(\ref{a2})

\begin{eqnarray}\label{a3}
(p_\mu \gamma^\mu -m_0)\delta \psi (p) = g F[\bar \psi_0   \psi_0]) \psi_0(p).
\end{eqnarray}
Multiplying the both sides of Eq.(\ref{a3}) by $(p_\mu \gamma^\mu +m_0)$ we obtain

\begin{eqnarray}\label{a4}
&&\delta \psi (p) = \frac{2m_0 gF [\bar \psi_0 \psi_0]}{p^2 - m_0^2} \psi_0(p)= \frac{2m_0 gF [\bar \psi_0 \psi_0]}{(\omega -\varepsilon(\mathbf p)+i\delta \operatorname{sgn}(\omega))(\omega +\varepsilon(\mathbf p)-i\delta \operatorname{sgn}(\omega))} \psi_0(p), 
\end{eqnarray}
where  $p=(\omega, \mathbf p)$ and $\varepsilon (\bb p)= \sqrt{\bb p^2 + m_0^2}$.

\begin{thebibliography}{9}

\bibitem{Gro}
 E. P. Gross,   Il Nuovo Cimento. \textbf { 20}, 454 (1961).
  \bibitem{Pit}
 L. P. Pitaevskii , Sov. Phys. JETP.  \textbf {13 }, 451 (1961).
\bibitem{Foo} 
C. J. Foot,  Atomic physics. Oxford University Press. pp. 231–24,  (2005)

\bibitem{Hug}
N. M. Hugenholtz; D. Pines,   Physical Review. {\textbf 116 }, 489 (1959).

 \bibitem{Bog}
 N.N.Bogolubov, J.Phys.USSR {\textbf 11},  23 (1947).
  \bibitem{Pit98} L.P.Pitaevskii,    Phys. Usp. {\textbf 41},  569 (1998)
  \bibitem{Oku} 
  A. Yu. Okulov , J. Phys. B: At. Mol. Opt. Phys. {\textbf41} 101001 (2008); Phys. Lett. {\textbf A 376},  650 (2012),
  \bibitem{LL9} E.M.Lifshitz, L.P.Pitaevskii.  Statistical Physics (Part 2, Theory of the Condensed State) , Pergamon Press,1980.
  \bibitem{Pes95}
M. E. Peskin, D. V. Schroeder, {\it An Introduction to Quantum
Field Theory}, Addison-Wesley Publishing Company, 1995.
\bibitem{LL4}
V. B. Berestetskii, E. M. Lifshitz, and L. P. Pitsevskii,
{\it Quantum Electrodynamics},
Pergamon Press, 1979.
\bibitem{Bar} J.Bardeen,  L. N.Cooper, J. R. Schrieffer,  Physical Review. {\textbf 106},  162 (1957).
\bibitem{Wil} K.G.Wilson, Phys. Rev. {\textbf D 10}, 2445 (1974).
\bibitem{Pol} A.M.Polyakov, Nuclear Physics {\textbf B120} , 429 (1977).
\bibitem{Gra}
\bibitem{20}
I.S.Gradshtain and I.M.Ryzhik, Tables of Integrals, Series and
Products, Academic Press, New York, 1980.

\end{thebibliography}
\end{document}